\documentclass[final,1p,times,twocolumn]{elsarticle}

\usepackage{amssymb}
\usepackage{amsmath}
\usepackage{subfigure}
\usepackage{graphicx}
\usepackage[colorlinks, linkcolor=blue, anchorcolor=blue,citecolor=blue, urlcolor=blue ]{hyperref}
\usepackage[T1]{fontenc}
\usepackage{slashed}
\usepackage[utf8]{inputenc}
\usepackage{xspace}
\usepackage{color}
\usepackage{array}
\usepackage{ulem,fancyvrb}
\usepackage{xcolor}

\journal{CPC}

\begin{document}

\begin{frontmatter}

\title{Using simulated Tianqin gravitational wave data and electromagnetic wave data to study the coincidence problem and Hubble tension problem}

\author[1,4]{Jia-Wei Zhang}
\author[1]{Jingwang Diao}
\author[2,4]{Yu Pan\corref{Pan}}
\author[3]{Ming-Yue Chen}
\author[4,4]{Jin Li\corref{Li}}

\address[1]{Department of Physics, Chongqing University of Science and Technology, Chongqing 401331, China}
\address[2]{School of Science, Chongqing University of Posts and Telecommunications, Chongqing 400065, China}
\address[3]{Department of physics, Chongqing University, Chongqing, 400044, China}
\address[4]{Department of Physics and Chongqing Key Laboratory for Strongly Coupled Physics, Chongqing University, Chongqing 401331, China}

\cortext[Pan]{Corresponding author, panyu@cqupt.edu.cn}
\cortext[Li]{Corresponding author, cqujinli1983@cqu.edu.cn}
\begin{abstract}
In this paper, we use electromagnetic wave data (H0LiCOW, $H(z)$, SNe) and gravitational wave data (Tianqin) to constrain the interacting dark energy (IDE) model and investigate the Hubble tension problem and coincidences problem. By combining these four kinds of data (Tianqin+H0LiCOW+SNe+$H(z)$), we obtained the parameter values at the confidence interval of $1\sigma$: $\Omega_m=0.36\pm0.18$, $\omega_x=-1.29^{+0.61}_{-0.23}$, $\xi=3.15^{+0.36}_{-1.1}$, and $H_0=70.04\pm0.42$ $kms^{-1}Mpc^{-1}$. According to our results, the best valve of $H_0$ show that the Hubble tension problem can be alleviated to some extent. In addition, the $\xi+3\omega_x = -0.72^{+2.19}_{-1.19}(1\sigma)$ of which the center value indicates the coincidence problem is slightly alleviated. However, the $\xi+3\omega_x = 0$ is still within the $1\sigma$ error range which indicates the $\Lambda$CDM model is still the model which is in best agreement with the observational data at present. Finally, we compare the constraint results of electromagnetic wave and gravitational wave on the model parameters and find that the constraint effect of electromagnetic wave data on model parameters is better than that of simulated Tianqin gravitational wave data.
\end{abstract}


\end{frontmatter}


\section{Introduction}

The observations of Type Ia supernovae (SNe) suggest that the universe is in a state of accelerated expansion  \cite{Riess1998,Weinberg2013,Scolnic2018,Perlmutter1999,Tegmark2004}. And subsequent observations of the cosmic microwave background (CMB) radiation and large-scale structure have further confirmed this view \cite{Spergel2003, Collaboration2020, Peebles2003, Bull2016, Bullock2017}. To this end, scientists have introduced a negative pressure energy as the driving force for the expansion of the universe, and called it dark energy. At the same time, many dark energy models have been proposed to study the nature of dark energy, of which the $\Lambda$ cold dark matter ($\Lambda$CDM) model is the most widely used and agrees with various cosmology observations. However, the $\Lambda$CDM model also faces some unresolved theoretical challenges \cite{Weinberg1989, Carroll1992}, namely the coincidence problem. This problem raises why the present epoch is so special that the density of dark energy is only on the same order of magnitude as the energy density of matter during this period. In addition, the measurement of $H_0$ in \cite{Riess+etal+2019} is $74.03\pm1.42$ $kms^{-1}Mpc^{-1}$ and the measurement of $H_0$ in \cite{Aghanim N+etal+2020} is $67.4\pm0.5$ $kms^{-1}Mpc^{-1}$, and the deviation between these two results is $4.4\sigma$.
It reflects the problem of measurement inconsistencies between the early and late universes. At present, there are two main directions to solve these problems: one is to use the new models \cite{Pan2020, Zlatev1999, Carroll1998A, Carroll1998B, Liddle1998, Caldwell2002, Carroll2003, Feng2005, Li2010, Yang2015, Wei2011, Wang2005}; the other is to find a more intuitive cosmological observation to obtain a more accurate Hubble constant, such as using gravitational wave data to constrain the cosmological model \cite{Abbott2017, Yang2017, Zhang2019A, Zhang2019B, Zheng2022}.

In this work, we use an interaction dark energy model, which considers a particular interaction. When we consider the proper interaction of dark energy with dark matter, dark energy can transform into dark matter, which in turn allows dark matter and dark energy to obtain some balance and thus alleviate the problem of coincidence.
And some observations also support this idea of interactions. For example, Bertolami et al. \cite{Bertolami2007, Bertolami2009, Delliou2007, Bertolami2008} pointed out that the Abelle group A586 indicates an interaction between dark energy and dark matter. In addition, Abdalla et al. \cite{Abdalla2009, Abdalla2010} found that optical, X-ray and weak lens data from the relaxation cluster were also used to obtain signals of energy exchange between the two. It is not enough to propose a cosmological model, but also to constrain it with the parameters of the observed model of various data. Chen et al. \cite{Chen2010} limited the model with data from supernovae (SNIa), CMB, and baryon acoustic oscillation (BAO), which showed that dark energy interaction with matter was necessary. Cao et al. \cite{Cao2013,Cao2015} used a combination of data such as Hubble parametric data and SNIa to constrain the model parameters, and the results show that the coincidence problem has not been alleviated, but the slight transfer of dark energy to dark matter is expected to be one of the solutions to alleviate the coincidence problem. In our previous work, based on the VLBI observation results of dense structures in 120 medium-luminance radio quasars (QSO), we combined the SNIa, CMB and BAO data \cite{Pan2020}. We also investigate the roles of strong gravitational lensing data in the study of dark energy by using measurements of the time-delay effect in 18 strong gravitational lensing systems \cite{Pan2016}.

In this article, the data we are using is simulated gravitational wave (GW) data from Tianqin. Gravitational wave sources from binary bodies and produced together provide a direct measurement of the luminosity distance. The Tianqin is a space-based GW detector developed by a team at Sun Yat-sen University in China, and its project concept has some unique features \cite{Shuai2020, Jianwei2021, Tod2005, Haitian2019, Ming2017, Hu2018}. The equilateral triangular arm is about $10^{5}$ km long, and the frequency sensitivity band of the detector overlaps with LISA at around $10^{4}$ Hz and DECIGO at around 0.1 Hz. Due to its short arm length and better sensitivity than LISA and Taiji in higher frequency bands, Tianqin fills the frequency gap between LISA and DECIGO. In addition, we also used $H(z)$, SNe and H0LiCOW data combined with Tianqin data to constrain the model.

The rest of the paper is organized as follows. In section \ref{sec:2}, we introduce the interaction dark energy model used in this article. The section \ref{sec:3} mainly introduces the gravitational wave data (Tianqin) and the electromagnetic wave data (H0LiCOW, $H(z)$, SNe) used in this paper. In section \ref{sec:4}, we use gravitational wave data and electromagnetic wave data to constrain the interacting dark energy model and analyze the results. At last, in section \ref{sect:5}, we make a summary of this paper.

\section{Model}\label{sec:2}

In order to describe dark energy, we choose a model of the interaction between dark energy density and dark matter density. In the flat FRW metric universe, we assume that dark energy and dark matter exchange energy through interaction $Q$ \cite{Pan2020}.

\begin{equation}
\dot{\rho_{X}}+3H\rho_{X}=-Q,
\end{equation}
\begin{equation}
\dot{\rho_{m}}+3H\rho_{m}=Q,
\end{equation}
where $\rho_{X}$ is the density of dark energy and $\rho_{m}$ is the density of matter. The $H(z)=\frac{\dot{a}}{a}$ means the Hubble parameter, where the $a=1/(1+z)$ and the $\dot{a}$ means derivative with respect to time. In addition, $\rho_{X}$ and $\rho_{m}$ have a phenomenological relationship $\frac{\rho_{X}}{\rho_{m}}\varpropto a^{\xi}$, where $\xi$ is the severity of the coincidence problem, and then we can get:
\begin{equation}
Q=-H\rho_{m}(\xi+3\omega_x)\Omega_{X},
\end{equation}
where $\omega_x$ is the equation of state of dark energy, $\Omega_{X}$ is the dark energy density parameter; it's expression is
 $\Omega_{X}=(1-\Omega_{m})/[1-\Omega_{m}+\Omega_{m}(1+z)^{\xi}]$.

When $\xi+3\omega=0(Q=0)$, it means that there is no interaction between dark energy and dark matter. When $\xi+3\omega>0(Q<0)$, dark matter is converted into dark energy and the coincidence problem is not alleviated. When $\xi+3\omega<0(Q>0)$, dark energy is converted into dark matter and the coincidence problem can be alleviated \cite{Cao2017,Cao2018,Diao2022}. The parameterized Friedman equation is expressed as:
\begin{equation}
E(z)^{2}=\frac{H^{2}}{H_{0}^{2}}=(1+z)^{3}[\Omega_{m}+(1-\Omega_{m})(1+z)^{-\xi}]^{-3\omega/\xi}.
\end{equation}

\section{Data}\label{sec:3}

\subsection{Tianqin}
It is believed that the supermassive binary black hole union is a good source of gravitational waves \cite{Wen-Fan Feng2019}, and the Tianqin detector can detect the supermassive black hole well  \cite{Y.-M2017,S.A.Hughes2001,A.Sesana2016,H.Di2018,S.Olmez2010}. For these reasons, we simulated 1000 sets of gravitational wave data from the merger of two black holes with a mass of $10^3$ solar masses. The redshift ($z$) of this data is $0\sim15$. It is important to note that when simulating the gravitational wave data, we adopt the flat $\Lambda$CDM with the matter density parameter $\Omega_m=0.286$ and the Hubble constant $H_0=69.6$ $kms^{-1}Mpc^{-1}$ with $1\%$ uncertainty \cite{Bennett14}.

Tianqin is a typical millihertz frequency gravitational wave observatory. Its main mission is to detect GWs from coalescing supermassive black hole binaries (SMBHBs), inspiral of stellar mass black hole binaries, Galactic ultra-binaries, extreme mass ratio inspirals (EMRIs), and stochastic GW background originating from primordial BHs or cosmic strings \cite{S.A.Hughes2001,A.Sesana2016,H.Di2018,S.Olmez2010}.
And the SMBHB mergers are the most powerful GW sources of them \cite{Wen-Fan Feng2019}.
That is why we simulated GW data set of SMBHBs, which have a mass of $10^3M_\odot$ and the redshift $z$ is chosen in the range of $0\sim15$.

We can obtain an absolute measure of the luminosity distance $D_{L}$ from the chirping GW signals of inspiral compact binary stars \cite{Schutz86}. The so-called chirp mass ($\mathcal{M}_c$) and the luminosity distance ($D_{L}$) decided the GW strain amplitude. The chirp mass can be obtained from the GW signal's phase position, so it is allowed to extract luminosity distance from the amplitude. The waveform function of GW we used is $h(f)$.
\begin{align}
h(f)=\mathcal{A}f^{-7/6}\exp[i(2\pi ft_0-\pi/4+2\psi(f/2)-\varphi_{(2.0)})],
\label{equa:hf}
\end{align}
where the Fourier amplitude $\mathcal{A}$ is
\begin{align}
\mathcal{A}=&~~\frac{1}{D_L}\sqrt{F_+^2(1+\cos^2(\iota))^2+4F_\times^2\cos^2(\iota)}\nonumber\\
            &~~\times \sqrt{5\pi/96}\pi^{-7/6}\mathcal{M}_c^{5/6},
\label{equa:A}
\end{align}
here the luminosity distance $D_{L}$ is very important for our purpose and can be expressed as a function of the redshift in the standard flat $\Lambda$CDM cosmological model.
\begin{equation}
D_{L}^{GW}=\frac{1+z}{H_{0}}\int^{z}_{0}\frac{dz'}{\sqrt{\Omega_{M}(1+z')^{3}+\Omega_{\lambda}}},
\end{equation}
 where $\Omega_{M}=0.286, \Omega_{\lambda}=0.714$ and $H_{0}=69.6 \text{km} \cdot\text{s}^{-1} \text{Mpc}^{-1}$. $M_c=(1+z)(m_{1}m_{2})^{3/5}/(m_{1}+m_{2})^{1/5}$ denotes the observed chirp mass. For Tianqin $F_{+}$, $F_{\times}$ and phase parameters can be found in \cite{Feng2019}.

Moreover, the one-sided power spectral density (PSD) of the detectors equivalent strain noise provided by Tianqin is as follows \cite{Luo2016,Hu2018}:

\begin{equation}
S_n(f)=\frac{S_x}{L_0^2}+\frac{4 S_a}{(2 \pi f)^4 L_0^2}\left(1+\frac{10^{-4} \mathrm{~Hz}}{f}\right),
\label{equa:Sn}
\end{equation}

where $L_0$ is the arm length and $L_0=1.73\times10^5 km$, the PSDs of the position noise is $S_x=10^{-24} m^2/Hz$ and the PSDs of the residual acceleration noise is $S_a=10^{-30} m^2 s^{-4}/Hz$.

Combining Eqs. \ref{equa:hf}, \ref{equa:A}, \ref{equa:Sn}, we can obtain the signal-to-noise ratio (SNR) of Tianqin
\begin{align}
\rho=\frac{M_{c}^{5 / 6}}{\sqrt{10} \pi^{2 / 3} D_{L}} \sqrt{\int_{f_{\text {in }}}^{f_{\text {fin }}} \frac{f^{-7 / 3}}{S_{n}(f)} \mathrm{d} f},
\end{align}
where $f_{fin} = min(f_{ISCO}, f_{end})$ with the GW frequency at the innermost stable circular orbit $f_{ISCO} =1/(6^{3/2}M\pi)$ Hz and the upper cutoff frequency for Tianqin $f_{end} = 1 Hz$, $f_{in} = max(f_{low}, f_{obs})$ with the lower cutoff frequency $f_{low} = 10^{-5}$ Hz and the initial observation frequency $f_{obs}=4.15\times10^{-5}(M_c/10^{6}M_\odot)^{-5/8}(T_{obs}/1 yr)^{-3/8}$ Hz. The observation time $T_{obs}$ is 3 months.

The inherent uncertainty of the measurement $\sigma^{inst}_{D_L}$ and the increased uncertainty due to weak lenses $\sigma^{lens}_{D_L}$ together make up the uncertainty of the luminosity distance. Therefore, we can calculate the total uncertainty of the measurement $D_L$ as \cite{Zhao2011,Sathyaprakash2010}
\begin{align}
\sigma_{D_{L}}^{GW}=&~~\sqrt{\left(\sigma_{D_{L}}^{inst}\right)^{2}+\left(\sigma_{D_{L}}^{lens}\right)^{2}} \\
&~~=\sqrt{\left(\frac{2D_{L}}{\rho}\right)^{2}+\left(0.05zD_{L}\right)^{2}},
\end{align}

In order to limit the model parameters, the $\chi^{2}$ minimum fitting method was adopted:
\begin{equation}
\chi^{2}=\sum_{i=1}^{1000}\frac{\left(D_{L}^{GW}(z)-D_{L}^{th}(z)\right)^{2}}{\left(\sigma_{D_{L}}^{GW}\right)^{2}},
\end{equation}

Where $D_{L}^{GW}$ is the luminosity distance observed by simulation, $\sigma_{D_{L}}^{GW}$ is the data error and the $D_{L}^{th}$ can be expressed as
\begin{equation}
D_{L}^{th}=\frac{1+z}{H_{0}}\int^{z}_{0}\frac{dz'}{\sqrt{(1+z)^{3}[\Omega_{m}+(1-\Omega_{m})(1+z)^{-\xi}]^{-3\omega/\xi}}}.
\end{equation}

\subsection{SNe and H(z)}
In addition, 580 groups of Ia SNe data are also used in this paper  \cite{Suzuki2012}. The data generally describe the brightness information of the supernova by distance modulus. The observed values in the data are expressed by the apparent magnitude m and absolute magnitude M as follows:
\begin{equation}
\mu_{\mathrm{obs}}=m-M,
\end{equation}
The theoretical value can also be obtained by the following formula:
\begin{equation}
\mu_{\mathrm{th}}=5 \log \left(d_{L} / \mathrm{Mps}\right)+25,
\end{equation}
The $\chi^{2}$ minimum fitting method related to distance modulus can be written as:
\begin{equation}
\chi_{\mathrm{SNIa}}^{2}=\sum_{i}^{580}\left(\mu_{\mathrm{obs}}-\mu_{\mathrm{th}}\right)^{2} / \sigma_{\mu, i}^{2},
\end{equation}
Where the $\sigma_{u, i}$ are the observation errors of supernovae.

Finally, the 31 Hubble parameter ($H(z)$) samples from the differential age method are also used \cite{J2017}.
And the $\chi^2$ minimum fitting method related to Hz parameter data set can be expressed as:
\begin{equation}
\chi_{H z}^{2}=\sum_{i=1}^{31}\left(\frac{H z_{t h}-H z_{o b s}}{\sigma_{H z}}\right)^{2}.
\end{equation}

\subsection{H0LiCOW}

The time delay phenomenon of the strong gravitational lensing in the late universe is an important cosmological probe that provides a method for measuring the Hubble constant ($H_{0}$)  \cite{Qi2022}. The time delay data we used came from H0LiCOW project ($H_{0}$ Lenses in COSMOGRAIL's Wellspring)  \cite{Kenneth2017, Rusu2017, Sluse2017, Bonvin2017}. In a strong gravitational lensing system, the source of stars will form multiple observation images under the action of the lens, and these images will take different paths to reach our detector, thus creating a time delay between the images. This data is sensitive to the Hubble constant $H_{0}$ and therefore can be used as a good constraint data \cite{Sluse2017}.

The time delay between images located at $\theta_{i}$ and $\theta_{j}$ can be expressed as
\begin{equation}
\tau(\theta_{i})-\tau(\theta_{j})=\frac{c\Delta t_{ij}}{D_{\Delta t}},
\end{equation}
where $\tau$ stands for dimensionless time of arrival, and $\tau=\frac{1}{2}\lvert\theta\lvert^{2}-\theta\cdot\beta$, the $\beta$ is the location of the source. $\Delta t_{ij}$ represents the measured value of time delay, $D_{\Delta t}$ is the called time delay distance, can be expressed as:
\begin{equation}
D_{\Delta t}=(1+z_{d})\frac{D_{d}D_{s}}{D_{ds}},
\end{equation}
where $D_{d}$ and $D_{s}$ are the angular diameter distance obtained when redshift is $z_{d}$ and $z_{s}$ respectively, and $D_{ds}$ is the angular diameter distance between the lens and the source.

In order to constrain cosmological parameters, we use the least squares fitting method to fit the parameters:
\begin{equation}
\chi^{2}_{D_{\Delta t}}=\sum_{i=1}^{18}\frac{[D^{th}_{\Delta t}(i)-D^{obs}_{\Delta t}(i)]^{2}}{\sigma(i)^{2}}.
\end{equation}
where $D_{\Delta t}^{th}$ is the time delay distance value theoretically existing in the cosmological model, $D_{\Delta t}^{obs}$ is the actual measured value, and its uncertainty is $\sigma(i)$.
\begin{figure}
   \centering
  \includegraphics[width=10cm, angle=0]{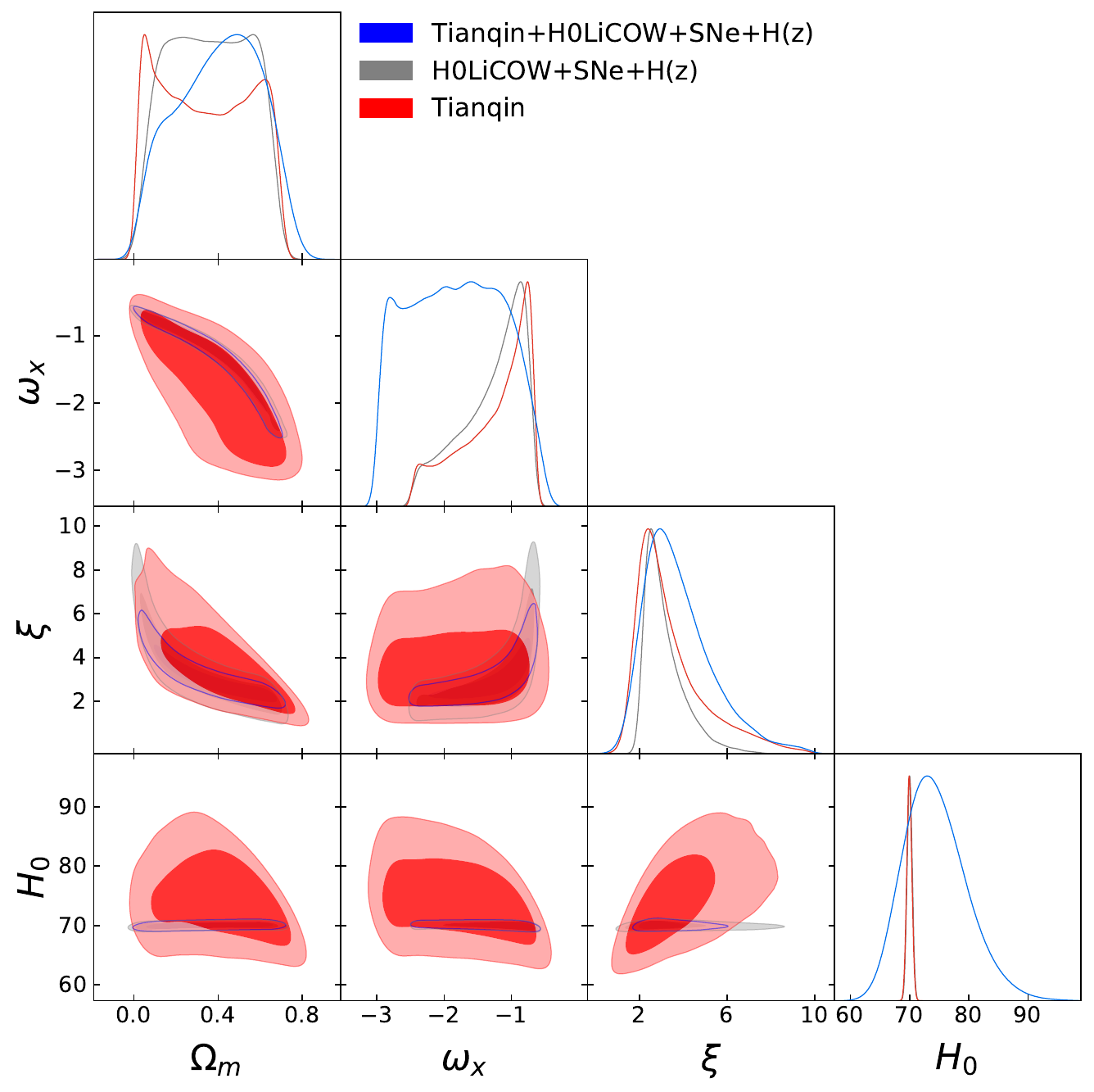}
   \caption{Contour map of Tianqin, H0LiCOW+SNe+$H(z)$ and Tianqin+H0LiCOW+SNe+$H(z)$ data combinations with constraints on model parameters ($\Omega_m$, $\omega_x$, $\xi$, $H_0$).}
   \label{p1}
\end{figure}

\section{Analysis of observation data constraint results}\label{sec:4}

In this paper, we used different combinations of four cosmological data, Tianqin, H0LiCOW, SNe and $H(z)$, to constrain the interaction model. They are Tianqin, $H(z)$+SNe, H0LiCOW, H0LiCOW+SNe+$H(z)$ and Tiqanqin+H0LiCOW+SNe+$H(z)$, respectively. And Markov Chain Monte Carlo (MCMC) algorithm and maximum likelihood method are used to limit the interacting dark energy model. For $\chi^{2}$ values of multiple sets of data, the following formula is used for combination:
\begin{equation}
\chi_{all}^{2}=\chi_{Tianqin}^{2}+\chi_{H0LiCOW}^{2}+\chi_{H(z)}^{2}+\chi_{SNe}^{2}.
\end{equation}

\begin{table}
\begin{center}
\caption{Parametric results obtained from different data constraint interaction models.}
\begin{tabular}{ccccc}
\hline
Data        &~~$\Omega_m$~~        &~~$\omega_x$~~      &~~$\xi$~~   &~~$H_0$~~    \\
\hline
Tianqin &$0.40_{-0.17}^{+0.25}(1\sigma)$ &$-1.83_{-0.67}^{+0.67}(1\sigma)$& $3.8_{-1.9}^{+0.89}(1\sigma)$& $74.4_{-6.0}^{+4.3}(1\sigma)$\\
\\
H(z)+SNe &$ 0.38_{-0.20}^{+0.20}(1\sigma)$& $-1.37_{-0.28}^{+0.68}(1\sigma)$& $3.39_{-1.6}^{+0.56}(1\sigma)$&$ 69.99_{-0.44}^{+0.44}(1\sigma)$\\
\\
H0LiCOW & $0.284_{-0.27}^{+0.067}(1\sigma)$& $-1.73_{-0.75}^{+0.21}(1\sigma)$& $4.5_{-1.6}^{+5.4}(1\sigma)$& $79.6_{-6.7}^{+4.8}(1\sigma)$\\
\\
H0LiCOW+SNe+H(z) & $0.34_{-0.21}^{+0.21}(1\sigma)$& $-1.26_{-0.22}^{+0.63}(1\sigma)$& $3.43_{-1.8}^{+0.54}(1\sigma)$& $69.97_{-0.42}^{+0.42}(1\sigma)$\\
\\
Tianqin+H0LiCOW+SNe+H(z) & $0.36_{-0.18}^{+0.18}(1\sigma)$& $-1.29_{-0.23}^{+0.61}(1\sigma)$& $3.15_{-1.1}^{+0.36}(1\sigma)$& $70.04_{-0.42}^{+0.42}(1\sigma)$\\
\hline
\end{tabular}
\label{tab1}
\end{center}
\end{table}

In Table \ref{tab1}, we show the constraint results of constraint model parameters($\Omega_{m}$, $\omega_{x}$, $\xi$, $H_{0}$) under five different data combinations. Table \ref{tab2} shows the parameter error estimates obtained under different data combinations, which are calculated as $\epsilon(p)=\sigma(p)/p$, where $p$ represents the parameter center value in the model and $\sigma(p)=\sqrt{(\sigma(p)^{2}_{upper}+\sigma(p)^{2}_{lower})/2}$.

As can be seen from Table \ref{tab1}, the Hubble constant ($H_0$) values given by the five data combinations with $1\sigma$ error are $\mathrm{H}_{0}=74.4_{-6.0}^{+4.3}kms^{-1}Mpc^{-1}$, $H_{0}=69.99_{-0.44}^{+0.44}kms^{-1}Mpc^{-1}$, $H_{0}=79.6_{-6.7}^{+4.8}kms^{-1}Mpc^{-1}$, $H_{0}=69.97_{-0.42}^{+0.42}kms^{-1}Mpc^{-1}$, $H_{0}=70.04_{-0.42}^{+0.42}kms^{-1}Mpc^{-1}$, respectively.
We can see that the value of Hubble constant $H_0$ given by ($H(z)$+SNe, H0LiCOW+SNe+$H(z)$, Tianqin+H0LiCOW+SNe+$H(z)$) at the confidence interval of $1\sigma$ are $H_{0}=69.99_{-0.44}^{+0.44}kms^{-1}Mpc^{-1}$, $H_{0}=69.97_{-0.42}^{+0.42}kms^{-1}Mpc^{-1}$ and $H_{0}=70.04_{-0.42}^{+0.42}kms^{-1}Mpc^{-1}$ which are smaller than $H_{0}=74.03_{-1.42}^{+1.42}kms^{-1}Mpc^{-1}$ given by the Hubble Space Telescope (HST) and lager than $H_{0}=67.4_{-0.5}^{+0.5}kms^{-1}Mpc^{-1}$ given by Plank2018, which indicates the conflict problem of $H_0$ has been alleviated to some extent. However, the center value of $H_0$ given by Tianqin and H0LiCOW data are lager than $H_{0}=74.03_{-1.42}^{+1.42}kms^{-1}Mpc^{-1}$ given by the HST, which indicates the Hubble tension problem has not been alleviated. In addition, we can see from Table \ref{tab2} that the constraint precision of $H_0$ given by the combination of electromagnetic wave data (H0LiCOW+SNe+$H(z)$) and gravitational wave data (Tianqin) is 0.006 and 0.081, respectively. This means that in this model, the electromagnetic wave data has a better constraint effect on $H_0$ than gravitational wave data. Furthermore, the constraint precision given by Tianqin+H0LiCOW+SNe+$H(z)$ is 0.006, which further indicates that the constraint effect of gravitational wave data on $H_0$ in this model is weak.

\begin{table}
\begin{center}
\caption{Constraint precision of the parameters ($\Omega_m, \omega_x, \xi, H_0$) obtained from different observational data.}
\begin{tabular}{ccccc}
\hline
Data& $\epsilon(\Omega_{m})$& $\epsilon(\omega_{x})$& $\epsilon(\xi)$& $\epsilon({H}_{0})$\\
\hline
Tianqin  & $0.625(1\sigma)$& $0.367(1\sigma)$& $0.5(1\sigma)$& $0.081(1\sigma)$\\
H(z)+SNe  & $0.526(1\sigma)$& $0.496(1\sigma)$& $0.472(1\sigma)$& $0.006(1\sigma)$\\
H0LiCOW  & $0.950(1\sigma)$& $0.433(1\sigma)$& $1.200(1\sigma)$& $0.084(1\sigma)$\\
H0LiCOW+SNe+H(z)  & $0.617(1\sigma)$& $0.500(1\sigma)$& $0.524(1\sigma)$& $0.006(1\sigma)$\\
Tianqin+H0LiCOW+SNe+H(z)  & $0.5(1\sigma)$& $0.473(1\sigma)$& $0.349(1\sigma)$& $0.006(1\sigma)$\\
\hline
\end{tabular}
\label{tab2}
\end{center}
\end{table}

In addition, other parameters besides Hubble's constant are given in Table \ref{tab1}, by which we can analyze the coincidence problem. In the previous content, we have talked that when $\xi+3\omega_x=0$, it means that there is no interaction, when $\xi+3\omega_x>0$, it means that dark matter is transitioned to dark energy and the coincidence problem has not been alleviated, when $\xi+3\omega_x<0$, dark energy is transitioned to dark matter and the coincidence problem can be alleviated. From Table \ref{tab1}, we can know that the $\xi+3\omega_x=-1.69^{+2.9}_{-3.91}$(Tianqin), $\xi+3\omega_x=-0.35^{+2.43}_{-2.46}$(H0LiCOW+SNe+$H(z)$) and $\xi+3\omega_x=-0.72^{+2.19}_{-1.79}$(Tianqin+H0LiCOW+SNe+$H(z)$). We can clearly see that the three data sets (Tianqin, H0LiCOW+SNe+$H(z)$, Tianqin+H0LiCOW+SNe+$H(z)$) each give a central value of $\xi+3\omega_x$ less than 0, which means that the coincidence problem is alleviated, but the $\xi+3\omega_x=0$ still within the $1\sigma$ error range. In addition, the result of parameter $\Omega_{m}$ obtained by the five data constraint models is consistent with that obtained by the CMB data from Planck satellite($\Omega_{m}=0.31\pm0.017(1\sigma)$) \cite{Ade2014} within the error range of $1\sigma$.

\section{Conclusion}
\label{sect:5}
In this paper, we used 31 groups of Hubble parameter observation data, 6 groups of H0LiCOW data, 580 groups of Type Ia supernova observation data and 1000 groups of Tianqin simulation data to constrain the parameters of the interacting dark energy model. Using Python programming to calculate data related formulas and MCMC algorithm program implementation. MCMC algorithm was used to calculate the $\chi^{2}$ test value of the observed data. The constraint results of the observational data on the model parameters and the circle graphs are shown in Table \ref{tab1} and Figure \ref{p1}, respectively.

(1) The value of Hubble constant $H_0$ given by ($H(z)$+SNe, H0LiCOW+SNe+$H(z)$, Tianqin+H0LiCOW+SNe+$H(z)$) at the confidence interval of $1\sigma$ are $H_{0}=69.99_{-0.44}^{+0.44}kms^{-1}Mpc^{-1}$, $H_{0}=69.97_{-0.42}^{+0.42}kms^{-1}Mpc^{-1}$ and $H_{0}=70.04_{-0.42}^{+0.42}kms^{-1}Mpc^{-1}$, respectively. Which indicates the conflict problem of $H_0$ has been alleviated to some extent.

(2) Comparing the constraint precision of the Hubble constant ($H_0$), we find that the electromagnetic wave data (H0LiCOW+SNe+$H(z)$) is better than the gravitational wave data for the model parameter $H_0$ in this interacting dark energy model.

(3) Based on the constraint results of the different observational data on the parameters $\xi$ and $\omega_{x}$, we can see that the $\xi+3\omega_x=-1.69^{+2.9}_{-3.91}$(Tianqin), $\xi+3\omega_x=-0.35^{+2.43}_{-2.46}$(H0LiCOW+SNe+$H(z)$) and $\xi+3\omega_x=-0.72^{+2.19}_{-1.79}$(Tianqin+H0LiCOW+SNe+$H(z)$). Which means that both electromagnetic wave data and gravitational wave data can alleviate the coincidence problem to some extent in this model. However, the $\xi+3\omega_x=0$ is still within the $1\sigma$ error range, which indicates that the $\Lambda$CDM model is still the model which is in best agreement with the observational data at present.

Finally, since the Tianqin Gravitational wave detector has not detected the GWs generated by high-redshift supermassive black hole binaries (SMBHBs) at present, we use the GW data with uniform distribution of redshift for simulation. In addition, the fact is that the H0LiCOW and Tianqin data have a weak constraint effect on the model parameters, which may be caused by the small H0LiCOW data and the unreal gravitational wave data. Therefor, we expect to detect more real gravitational wave data and strong gravitational lensing data in the future to help us further investigate the coincidence problem and Hubble tension problem.
\section*{Acknowledgments}

This work is supported in part by the National Natural Science Foundation of China (Grant Nos. 12105032, 12147102), Jia-Wei Zhang was supported by the Natural Science Foundation of China under Grants No. 12275036, the Natural Science Foundation of Chongqing under Grants No. cstc2021jcyj-msxmX0681 and No.cstc2018jcyjAX0713, No.cstc2021jcyj-msxmX0553. The Science and Technology Research Program of Chongqing Municipal Education Commission under Grant No.KJQN202001541, and the Research Foundation of Chongqing University of Science and Technology under Grant No.CK2016Z03. This work was also supported by the Graduate Research and Innovation Foundation of Chongqing, China (Grant No. CYS21327). Authors Jia-Wei Zhang and Jing-Wang Diao contributed equally to this article. Authors Jia-Wei Zhang and Jing-Wang Diao are co-first authors of the article.




\end{document}